# High spatial frequency periodic structures induced on ferric ion-doped Polyvinyl Pyrrolidone film by femtosecond laser pulses

Chen Lai, Guocai Liao, Yunxia Wang, Qiang Li, and Lijun Wu[*]

*Guangdong Provincial Key Laboratory of Nanophotonic Functional Materials and Devices, School of Information and Optoelectronic Science and Engineering, South China Normal University, Guangzhou 510006, P.R. China*

[*] ljwu@scnu.edu.cn

**Abstract:** Utilizing continues-wave or pulsed laser to induce nano-structures on various material surfaces is one significant method in nano-fabrication technology. In this report, we investigate the formation of high spatial frequency periodic structures on Polyvinyl Pyrrolidone (PVP) film by a linearly polarized femtosecond laser. Ferric (Fe) ions are introduced into the film to improve the photosensitivity. Regular nano-gratings with spatial periods at the range of 60-100nm, which are about one tenth of the irradiating wavelength, can be induced. The period direction of the nano-gratings is perpendicular to the polarization of the femtosecond laser. By tuning the laser energy and scanning speed, we find that the nano-gratings can be formed in a wide range of experimental parameters. As high laser energy can excite not only metals, but also semiconductors and polymers, we believe the formation of the nano-gratings is due to the interaction between the incident femtosecond laser and surface plasmons. The laser processable PVP-based materials and the induced nano-gratings will have potential applications in biophotonics and nanophotonics.

## 1. Introduction

Laser-induced periodic surface structures (LIPSSs) which also termed as ripples on various materials including metals, semiconductors and dielectrics have been widely investigated by using continuous-wave (CW) or pulsed lasers [1-5].The ripples with period $\Lambda$ approximately equal to the laser wavelength are called low spatial frequency LIPSSs (LSFL), whose orientations are always perpendicular to the laser beam polarization [6-8]. On the other hand, the ripples with $\Lambda$ much smaller than the laser wavelength are referred to as high spatial frequency LIPSSs (HSFL). Their orientations are often perpendicular [9] and sometimes parallel to the laser beam polarization [10]. When ultra-short laser pulses such as femtosecond are applied, the size of LIPSSs can be much smaller than the irradiating wavelength or even below one tenth of it [11, 12]. Thus ultra-short laser pulses offer a potential tool for producing surface nano-structures [13, 14].

The formation of LIPSSs is a complex process. Miscellaneous interactions between light and matter such as laser-induced carrier excitation or carrier thermalization are involved [15-

17]. Several models have been developed to explain the underlying physics [3, 18-20]. Among them, the surface scattered wave model is the most popular and widely accepted one. It contributes the formation of LSFL to the interference between the incident wave and the surface scattered waves caused by the surface roughness of the sample. The period of LSFL is given by $\lambda/(1\pm\sin\theta)$, where $\lambda$ is the irradiating wavelength and $\theta$ the angle of incidence.

However, the period of HSFL cannot be described by the aforementioned equation. More factors such as excitation of surface plasmon polaritons (SPPs) or grating-assisted SP-laser coupling have been taken into account [21, 22]. Under irradiation of ultra-short laser pulses, the sample surface can be highly excited regardless of the type of the material [22, 23]. The optical properties of the sample are modified and thus SPPs can even be excited in non-metallic materials. Other possible physical mechanisms such as self-organization [24], second harmonic generation [25], and Coulomb explosion [26] have also been proposed to interpret the formation of HSFL. However, there is still no comprehensive mechanism to explain the actual physical process until now.

For HSFL, most of the investigations have been focused on semiconductors or metals [27-30]. The period of ripples depends strongly on the type of materials with different refractive indexes or band gaps [31]. Polymer is another very important class of materials which is significant in the process of chemistry and biology. However, to the best of our knowledge, there are very few reports on the formation of HSFL on polymer-based materials. Polyvinyl Pyrrolidone (PVP) is one of the most popular polymer widely utilized in medical and food industries because of its excellent solubility, hypotoxicity, chemicalstability, film-forming and biodegradable properties. The absorption efficiency of PVP is very low at visible. When it is irradiated by an ultraviolet (UV) laser, ripples could be formed with a period similar to the laser wavelength and parallel to the laser polarization direction [32]. Nonetheless, UV light is harmful to biological cells thus it is not suitable for PVP processing in biology applications. On the other hand, some groups have added PVP into gold salt [33], platinum/palladium salt [34] or silver salt [35] solutions to investigate its influence when the materials are irradiated by pulsed laser. The interaction between the materials and laser is a photoreduction process through which the metallic ions can be reduced into atomic state and metallic nanoparticles can be formed finally. PVP functions as a supportive or capping material in these cases. No LIPSSs have been found in these material systems [33-35].

In this paper, we investigate the interaction between the PVP-based polymer and femtosecond pulsed laser. $Fe(NO_3)_3$ is added into the PVP solution to increase its photosensitivity. We find that HSFL (also named as nano-gratings here due to their high regularity) can be formed on the Fe ion-doped PVP film with orientation being always perpendicular to the polarization direction of the laser beam. The period of the nano-gratings is found to be between 60-100nm, which breaks the optical diffraction limit and could not be obtained by conventional photon lithography. Since the optical properties of materials could be modified by ultra-short laser pulses, we believe that the formation of these nano-gratings is due to the excitation of SPPs [36], which is similar to the formation mechanism in metals and semiconductors [23, 37]. By investigating the influence of the laser energy and the scanning speed of the sample on the formation of nano-gratings, we find that they can be produced in a broad range of experiment parameters. These nano-gratings formed in PVP-based materials can find applications in biophotonics, nanophotonics and optoelectronics devices.

## 2. Experimental setup

PVP（Shanghai EKEAR）and $Fe(NO_3)_3$ (Shanghai Crystal Pure) were mixed in deionized water and stirred for 20minutes. Then they were spin-coated on cover glasses and dried at 100℃ for 10minutes. After written by the pulsed laser, the Fe ion-doped PVP film was soaked in ethyl alcohol for two hours and the unexcited region was then washed out by deionized water. The thickness of the film was controlled to be between 60-80nm. The laser direct-writing system is similar to that described in the reference [38]. Shortly, laser pulses at a wavelength of 800nm with pulse duration 130fs were delivered from a mode-locked Ti: sapphire oscillator at 76MHz repetition rate (Mira 900, Coherent). An attenuator and a half wave plate were used to control the energy and the polarization direction of the laser beam. The polarized laser beam was focused normally on the sample by a 100x oil-immersion objective (NA=1.4, Zeiss). The diameter of the focused spot is about 1μm. The sample was fixed on a computer-controlled three-dimensional high-precision nanopositioning translation stage (P-563, PI). During the writing process, the sample was displaced by the nanopositioning stage while the laser beam was stationary. In order to investigate the influence of the experimental conditions on the formation of nano-gratings, the transmitting speed of the sample was varied from 0.2μm/s to 20μm/s and the laser energy from $51mJ/\mu m^2$ to $127mJ/\mu m^2$. All experiments were carried out in air at room temperature. The surface morphology of the written samples was examined by a scanning electron microscope (SEM) (Ultra 55, Zeiss).

## 3. Results and Discussions

As shown in Fig. 1, the absorption of the PVP solution is very low from 350nm to 900nm, as was mentioned in the introduction part. It can be significantly increased by introducing Fe$(NO_3)_3$ into the PVP solution at around 425nm. Not surprisingly, higher Fe concentration leads to larger absorption. For pattern writing, a molar ratio of 1:1 between PVP and Fe$(NO_3)_3$ was utilized in the experiments.

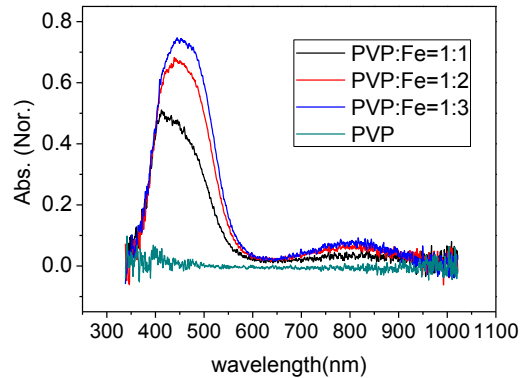

Fig. 1. The absorption spectra (normalized to water) for PVP and Fe ion-doped PVP solutions.

Fig. 2(a) exhibits a typical SEM image for a written line on the Fe ion-doped PVP film. As can be seen, there are three different regions on the sample: unexcited area, solidified area and nano-grating area. It's known that the material response under different energy irradiation is different. As the nano-gratings were always found to be in the center of the laser irradiation track in our experiments, we can deduce that the energy to induce nano-gratings is higher than that to induce solidification. Therefore, we can divide the Gaussian laser beam into three different regions as shown in Fig. 2(b). When the energy of the laser beam at focus approaches the first threshold level (T1), the Fe ion-doped PVP film would be cross-linked and solidified. If the energy exceeds the second threshold level (T2), nano-gratings would be formed. The period of the nano-gratings was found to be between 60-100nm in our experiments, which is about 1/10 of the irradiation wavelength. Such small value breaks the diffraction limit and cannot be obtained by conventional photo-lithography technique. The width and the length of the nano-grating are approximately equal to 40nm and 600nm respectively. The depth of the nano-grating is in the range of 20-40nm, depending on the thickness of the film. These parameters are reproducible and remain similar under different experimental parameters. If we further increase the laser energy to surpass the third threshold level (T3), the Fe ion-doped PVP film would be ablated and nothing would be left on the sample. One interesting thing happened in our experiments was that we have not found low spatial frequency LIPSSs in our material system.

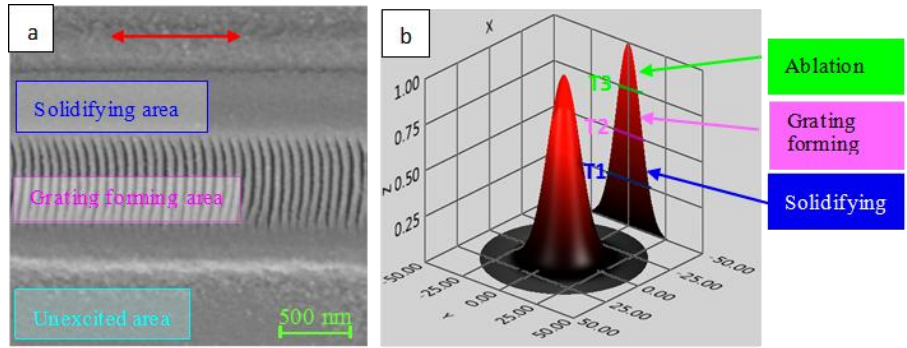

Fig. 2. (a) A typical SEM image for a written line showing different response regions which correspond to different energy level of the Gaussian laser beam. The scanning speed and laser energy are 2μm/s and 75mJ/μm$^2$ respectively. The red double-direction arrow plots the polarization direction of the laser beam. (b) A schematic cross section of the Gaussian laser beam. T1, T2, T3 represent the three threshold levels respectively.

Normally, the direction of the induced nano-gratings with either HSFL or LSFL is dependent on the polarization of the irradiation laser [31]. Here we vary the scanning and polarization directions to investigate their influences. Fig. 3(a) demonstrates a circle drawn by the pulsed laser. The polarization was along the vertical direction. As can be seen, the direction of the nano-gratings remain horizontal across the whole circle, meaning that it is always perpendicular to the polarization of the irradiation regardless of the laser scanning direction. The disconnection on the written circle indicated by a blue dotted square is caused by the misregistration of the nanopositioning translation stage between the starting and end points of the scanning. Fig. 3(b) shows a non-orthogonal cross written by the horizontally polarized laser beam. The direction of the nano-gratings confirms that it is perpendicular to the polarization of the irradiation laser beam. Because of the asymmetry of the focus spot, the width of the solidified area besides the nano-grating area is found to be asymmetric as well, which can be observed in Fig. 3(a) and Fig. 3(b).

Obviously, the scanning speed and the power of the laser are correlated when we consider the energy acts on the sample in the laser writing experiments. Lager scanning speed corresponds to less pulses and thus lower laser energy accumulated in the material within a certain period of time. Now we investigate the influence of the scanning speed as well as the laser energy on the formation of HSFL respecvively.

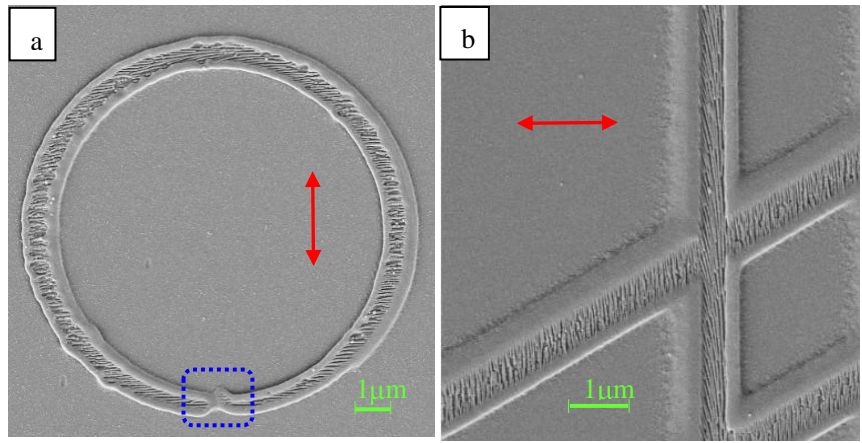

Fig. 3. SEM images show the relationship between the laser polarization direction and the orientation of the nano-gratings. (a) A circle written by a vertically polarized laser beam at a scanning speed of 2μm/s with the laser energy 83mJ/μm$^2$. (b) A non-orthogonal cross written by a horizontally polarized laser beam at a scanning speed of 2μm/s with the laser energy 7mJ/μm$^2$. The red double-direction arrows plot the polarization direction of the laser beam.

At first, we fixed the laser energy to be 76mJ/μm$^2$ and tuned the scanning speed to investigate the evolution of the surface nano-strucutres. As demonstrated in Fig. 4, the evolution of the surface morphology can be divided into four stages with different scanning speed. At a scanning speed of 8μm/s, the film starts to be cross-linked (refer to Fig. 4(a)). This corresponds to the solidified region as shown in Fig. 2(b). A point defect circled by red dotted-line can be observed on the right hand side of the picture. With a decrease of the scanning speed, more defects are formed (refer to Fig. 4(b) and 4(c)). At the same time, the defects grow transversely and form nano-gratings with more pulses accumulated in the materials (refer to Fig. 4(d) and 4(e)). If too much energy is accumulated in the materials within a period of time, the energy threshold T3 is to be reached and the materials would be ablated. The last transverse length of the nano-gratings is dependent on the diameter of the focused laser spot. In our experiments, the size of the focus spot is ~1μm. The length of the nano-gratings was found to be ~600nm.

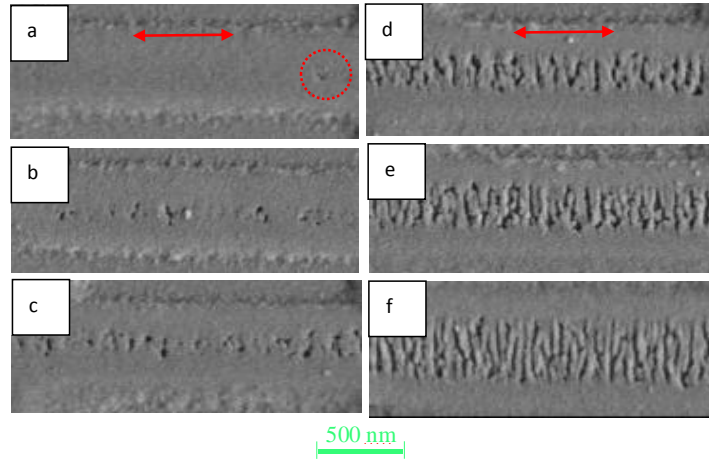

Fig. 4. SEM images of samples under laser energy of 76mJ/μm$^2$ with different scanning speed (a) 8μm/s, (b) 4μm/s, (c) 2μm/s, (d) 1μm/s, (e) 0.5μm/s, (f) 0.2μm/s. The red double-direction arrows plot the polarization direction of the laser beam.

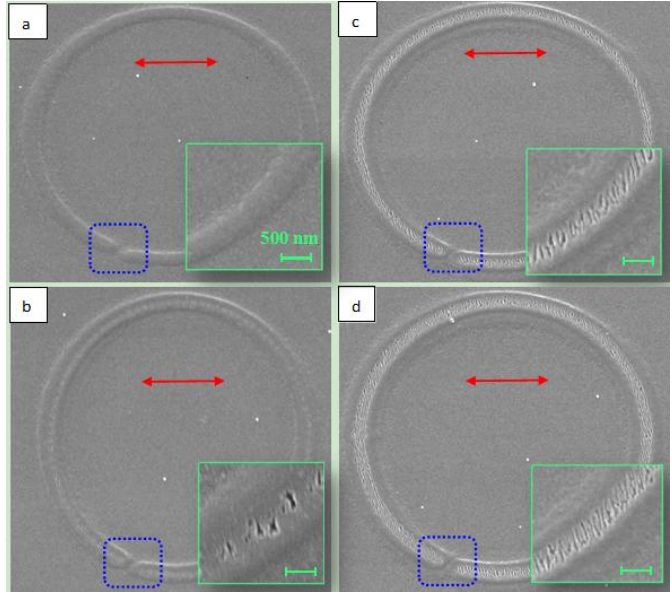

Fig. 5. SEM images of samples under laser energy of 70mJ/μm$^2$ with different scanning speed (a) 6μm/s, (b) 4μm/s, (c) 2μm/s, (d) 1μm/s respectively. The right insets in each image are the magnified parts of the written circle to demonstrate the details of the nano-structures. The unconnected parts on the circles positioned by blue dotted squares are caused by the misregistration of the nanopositioning translation stage between the starting and end points of the scanning. The red double-direction arrows plot the polarization direction of the laser beam. The size of the green scale bars is 500nm.

When we changed the scanning track of the laser beam from a straight line to a circle, as shown in Fig. 5, the response trend of the material is similar. In other words, more nano-holes are induced and grow transversely into nano-gratings simultaneously when more laser pulses are accumulated in the materials (refer to Fig. 5(b-d)). Therefore, the processes of forming nano-gratings are similar for different scanning patterns.

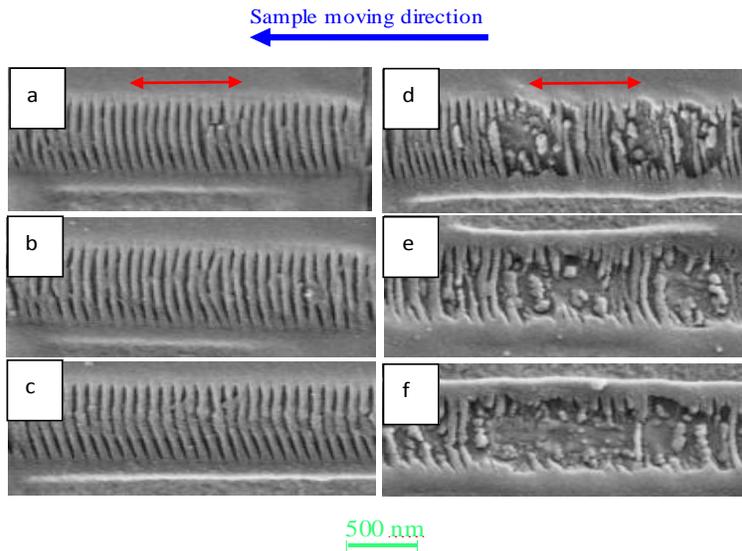

Fig. 6. SEM images of samples under laser energy of 83mJ/μm$^2$ with different scanning speed: (a) 0.5μm/s, (b) 1μm/s, (c) 2μm/s, (d) 4μm/s, (e) 8μm/s, (f) 12μm/s. The red double-direction arrows plot the polarization direction of the laser beam.

The lowest speed of our scanning system is limited to be 0.1μm/s. As the scanning speed and the energy accumulated in the materials are correlated, we increased the laser power to be 89mJ/μm$^2$, which can achieve similar effect as decreasing the scanning speed further. The results are shown in Fig. 6(a-f). Obviously, with an increase of the laser power, the influence of the scanning speed becomes different compared to the results shown in Fig. 4 and Fig. 5. The nano-gratings can be formed in a wide range of scanning speed. When the speed is faster than 2μm/s, some defects start to appear at the center of the nano-gratings. With a further increase, the defects dominate the nano-grating area and there are almost no nano-gratings at last, as shown in Fig. 6(f). This may be due to a strong thermal effect induced by high laser energy. When the laser power is within the nano-grating forming range, the Fe ion-doped PVP may be softened or melted, especially at the center of the laser beam. If the scanning speed is too fast, the softened center part of the cross-linked polymer may be drawn by the movement of the sample. This point can be confirmed by the bending direction of the nano-gratings. The results shown in Fig. 6(a-f) were obtained by moving the sample from right to left. As can be observed, all the nano-gratings bend to the moving direction of the sample. This phenomenon suggests that the formation of the nano-gratings requires the laser beam dwelt on the sample for a period time.

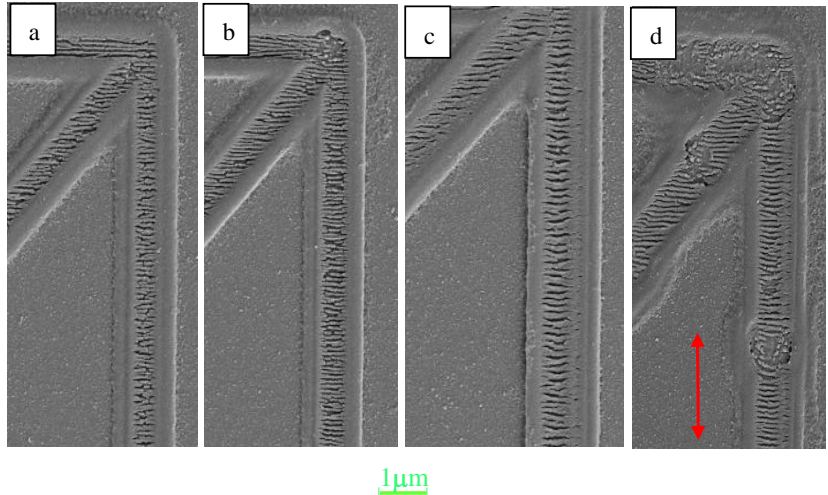

Fig. 7. SEM images of samples at a scanning speed of 1μm/s with laser energy: (a) 57mJ/μm$^2$, (b) 70mJ/μm$^2$, (c) 83mJ/μm$^2$, (d) 95mJ/μm$^2$. The red double-direction arrow plots the polarization direction of the laser beam.

On the other hand, the laser energy also influences the formation of the nano-gratings. Fig. 7 exhibits typical SEM images at a scanning speed of 0.5μm/s under different laser energy. As shown, well-ordered nano-gratings can be produced within a laser energy range of 57-83mJ/μm$^2$. When the laser energy is as high as 95mJ/μm$^2$, some parts of the nano-gratings were ablated as shown in Fig. 7(d).

The above experimental results show that, although the scanning speed of the sample and the laser beam energy influence the quality of the nano-gratings, they can be produced easily at a wide range of experimental parameters. Furthermore, we expect other reductives such as copper salt could also be utilized to enhance the absorption efficiency, which makes PVP processable by femtosecond laser at visible. Therefore, the method demonstrated here is highly flexible and has potential applications for generating nano-structures much smaller than the irradiating wavelength in nanophotonics and bio-nanophotonics.

## 4. Conclusions

To summarize, we have investigated the interaction between femtosecond pulsed laser and

PVP-based polymer film. It was found that the Fe ions doped into the PVP film can enhance its absorption efficiency and make it processable by femtosecond laser at visible. High spatial frequency nano-gratings with periods between 60-100nm can be induced on the PVP-based film which breaks the diffraction limit. The orientation of the nano-gratings was found to be perpendicular to the polarization of the laser beam. Although both the sample scanning speed and laser energy can influence their formation, the nano-gratings can be produced very easily in a wide range of experimental parameters. These nano-gratings produced in PVP-based materials will have potential applications in various biotechnology and nano-optoelectronics.

**Acknowledgments**

The authors acknowledge the financial support from National Natural Science Foundation of China (Grant No. 61378082) and the Project of High-level Professionals in the Universities of Guangdong Province.

**References and links**

1.  M. Birnbaum, "Semiconductor surface damage produced by ruby lasers," J. Appl. Phys. **36**(11), 3688-3689 (1965).
2.  P. M. Fauchet and A. E. Siegman, "Surface ripples on silicon and gallium arsenide under picosecond laser illumination," Appl. Phys. Lett. **40**(9), 824-826 (1982).
3.  J. E. Sipe, J. F. Young, J. S. Preston, and H. M. Van Driel, "Laser-induced periodic surface structure. I. Theory," Phys. Rev. B **27**(2), 1141-1154 (1983).
4.  J. S. Preston, H. M. van Driel, and J. E. Sipe, "Pattern formation during laser melting of silicon," Phys. Rev. B Condens. Matter. **40**(6), 3942-3954 (1989).
5.  S. E. Clark and D. C. Emmony, "Ultraviolet-laser-induced periodic surface structures," Phys. Rev. B Condens.Matter **40**(4), 2031-2041 (1989).
6.  J. Bonse, A.M. Munz, and H. Sturm, "Structure formation on the surface of indium phosphide irradiated by femtosecond laser pulses," J. Appl. Phys. **97**(1), 013538(2005).
7.  A. Y. Vorobyev, V. S. Makin, and C. Guo, "Periodic ordering of random surface nanostructures induced by femtosecond laser pulses on metals," J. Appl. Phys. **101**(3), 034903 (2007).
8.  J. Bonse, A. Rosenfeld, and J. Krüger, "On the role of surface plasmon polaritons in the formation of laser-induced periodic surface structures upon irradiation of silicon by femtosecond laser pulses," J. Appl. Phys. **106**(10), 104910 (2009).
9.  M. Rohloff, S. K. Das, S. Höhm, R. Grunwald, A. Rosenfeld, J. Krüger, and J. Bonse, "Formation of laser-induced periodic surface structures on fused silica upon multiple cross-polarized double-femtosecond-laser-pulse irradiation sequences," J. Appl. Phys. **110**(1), 014910(2011).
10. T. Q. Jia, H. X. Chen, M. Huang, F. L. Zhao, J. R. Qiu, R. X. Li , and H. Kuroda, " Formation of nanogratings on the surface of a ZnSe crystal irradiated by femtosecond laser pulses," Phys. Rev. B **72**(12),125429(2005).
11. J. Bonse, S. Höhm, A. Rosenfeld, and J. Krüger, "Sub-100-nm laser-induced periodic surface structures upon irradiation of titanium by Ti: sapphire femtosecond laser pulses in air," Applied Physics. A **110**(3), 547 551(2013).
12. J. Bonse, J. Krüger, S. Höhm, and A. Rosenfeld, "Femtosecond laser-induced periodic surface structures," J. Laser Appl. **24**(4), 042006(2012).
13. M. Groenendijk and J. Meijer, "Microstructuring using femtosecond pulsed laser ablation," J. Laser Appl. **18**(3), 227-235(2006).
14. R. G. Gattass and E. Mazur, "Femtosecond laser micromachining in transparent materials," Nature Photonics 2, 219-225 (2008).
15. A. Tien, S. Backus, H. Kapteyn, M. Murnane, and G. Mourou, "Short-pulse laser damage in transparent materials as a function of pulse duration", Phys. Rev. Lett. 82, 3883-3886 (1999).
16. L. Sudrie, A. Couairon, M. Franco, B. Lamouroux, B. Prade, S. Tzortzakis, and A. Mysyrowicz, "Femtosecond laser-induced damage and filamentary propagation in fused silica," Phys. Rev. Lett. **89**(18), 186601 (2002).
17. C. B. Schaffer, A. Brodeur, and E. Mazur, "Laser-induced breakdown and damage in bulk transparent materials induced by tightly-focused femtosecond laser pulses," Meas. Sci. Technol. 12, 1784 (2001).
18. P. M. Fauchet and A. E. Siegman, "Surface ripples on silicon and gallium arsenide under picosecond laser Illumination," Appl. Phys. Lett. **40**(9), 824-826(1982).
19. A. E. Siegman and P. M. Faucher, "Stimulated Wood's Anomalies on Laser-Illuminated Surfaces," IEEE J. Quantum Electron. **22**(8), 1384–1403 (1986).
20. M. Oron and G. So, "New experimental evidence of the periodic surface structure in laser annealing," Appl. Phys. Lett. **35**(10), 782-784(1979).

21. D. J. Ehrlich, S. R. J. Brueck, and J. Y. Tsao, "Time-resolved measurements of stimulated surface polariton wave scattering and grating formation in pulsed-laser-annealed germanium," Appl. Phys. Lett. **41**(7), 630-632 (1982).
22. M. Huang, F. Zhao, Y. Cheng, N. Xu and Z. Xu, "Origin of laser-induced near-subwavelength ripples: interference between surface plasmons and incident laser," Acs Nano. **3**(12), 4062-4070 (2009).
23. M. Huang, F. Zhao, Y. Cheng, N. Xu, and Z. Xu, "Mechanisms of ultrafast laser-induced deep-subwavelength gratings on graphite and diamond," Phys. Rev. B **79**(12), 125436(2009).
24. F. Costache, M. Henyk, and J. Reif, "Modification of dielectric surfaces with ultra-short laser pulses," Appl. Surf. Sci. 186, 352-357 (2002).
25. T. Q. Jia, H. X. Chen, M. Huang, F. L. Zhao, J. R. Qiu, R. X. Li, and H. Kuroda, "Formation of nanogratings on the surface of a ZnSe crystal irradiated by femtosecond laser pulses," Phys. Rev. B **72**(12), 125429(2005).
26. Y. Dong and P. Molian, "Coulomb explosion-induced formation of highly oriented nanoparticles on thin films of 3C–SiC by the femtosecond pulsed laser," Appl. Phys. Lett. **84**(1), 10-12 (2004).
27. R. Le Harzic, H. Schuck, D. Sauer, T. Anhut, I. Riemann, and K. König, "Sub-100 nm nanostructuring of silicon by ultrashort laser pulses," Opt. Express **13**(17), 6651-6656(2005).
28. S. Höhm, A. Rosenfeld, J. Krüger, and J. Bonse, "Femtosecond laser-induced periodic surface structures on silica," J. Appl. Phys. **112**(1), 014901(2012).
29. L. Qi, K. Nishii, and Y. Namba, "Regular subwavelength surface structures induced by femtosecond laser pulses on stainless steel," Opt. Lett. **34**(12), 1846-1848(2009).
30. X. Jia, T. Q. Jia, Y. Zhang, P. X. Xiong, D. H. Feng, Z. R. Sun, and Z. Z. Xu, "Periodic nanoripples in the surface and subsurface layers in ZnO irradiated by femtosecond laser pulses," Opt. Lett. **35**(8), 1248-1250(2010).
31. J. Bonse, J. Krüger, S. Höhm, and A. Rosenfeld, "Femtosecond laser-induced periodic surface structures," J. Laser Appl. **24**(4), 042006 (2012).
32. S. Pérez, E. Rebollar, M. Oujja, M. Martín, and M. Castillejo, "Laser-induced periodic surface structuring of biopolymers," Appl. Phys., A Mater. Sci. Process.**110**(3), 683–690(2013).
33. W. E. Lu,Y. L. Zhang, M. L. Zheng, Y. P. Jia, J. Liu, X. Z. Dong, and X. M. Duan, "Femtosecond direct laser writing of gold nanostructures by ionic liquid assisted multiphoton photoreduction," Opt. Mater. Express **3**(10), 1660-1673(2013).
34. L. D.Zarzar, B. S. Swartzentruber, J. C. Harper, D. R. Dunphy, C. J.Brinker, J. Aizenberg, and B. Kaehr "Multiphoton lithography of nanocrystalline platinum and palladium for site-specific catalysis in 3D Microenvironments," J. Am. Chem. Soc. **134**(9), 4007-4010(2012).
35. S. Maruo and T. Saeki, "Femtosecond laser direct writing of metallic microstructures by photoreduction of silver nitrate in a polymer matrix," Opt. Express **16**(2), 1174–1179 (2008).
36. S. K. Das, H. Messaoudi, A. Debroy, E. McGlynn, and R. Grunwald, "Multiphoton excitation of surface plasmon-polaritons and scaling of nanoripple formation in large bandgap materials," Opt. Mater. Express. **3**(10), 1705-1715(2013).
37. J. W. Yao, C. Y. Zhang, H. Y. Liu, Q. F. Dai, L. J. Wu, S. Lan, A. V. Gopal, V. A. Trofimov, and T. M. Lysak, "High spatial frequency periodic structures induced on metal surface by femtosecond laser pulses," Opt. Express **20**(2), 905–911 (2012).
38. P. Liu, J. Lan, J. Hu, S. Zhang, and Y. Lu, "Self-organizing microstructures orientation control in femtosecond laser patterning on silicon surface," Opt. Express, **22**(14), 16669-16675(2014).